\documentclass[aps,prd,twocolumn,showpacs,superscriptaddress,preprintnumbers]{revtex4-1}
\usepackage[colorlinks=true, pdfstartview=FitV, linkcolor=red,citecolor=blue, urlcolor=blue,
bookmarks=true, bookmarksnumbered=true, breaklinks]{hyperref}
\usepackage[dvipdfmx]{graphicx}
\usepackage{amsmath,amssymb,tabularx,ulem,mathrsfs,bm}
\begin{document}

\title{Cigar-shaped quarkonia under strong magnetic field}

\author{Kei Suzuki}
\affiliation{Theoretical Research Division, Nishina Center, RIKEN, Wako, Saitama, 351-0198, Japan}

\author{Tetsuya Yoshida}
\affiliation{Theoretical Research Division, Nishina Center, RIKEN, Wako, Saitama, 351-0198, Japan}
\affiliation{Department of Physics, Tokyo Institute of Technology, Meguro, Tokyo, 152-8551, Japan}

\date{\today}
\preprint{RIKEN-QHP-214}

\begin{abstract}
Heavy quarkonia in a homogeneous magnetic field are analyzed by using a potential model with constituent quarks.
To obtain anisotropic wave functions and corresponding eigenvalues, the cylindrical Gaussian expansion method is applied, where the anisotropic wave functions are expanded by a Gaussian basis in the cylindrical coordinates.
Deformation of the wave functions and the mass shifts of the $S$-wave heavy quarkonia ($\eta_c$, $J/\psi$, $\eta_c(2S)$, $\psi(2S)$ and bottomonia) are examined for the wide range of external magnetic field.
The spatial structure of the wave functions changes drastically as adjacent energy levels cross each other.
Possible observables in heavy-ion collision experiments and future lattice QCD simulations are also discussed.
\end{abstract}
\maketitle

\section{Introduction}
The deformation of nuclei, as shown first by Rainwater, Bohr and Mottelson, has been studied for a long time in nuclear physics, where a nucleus exhibits cigar-shaped (prolate) or disk-shaped (oblate) deformation and a complicated energy-level structure known as a Nilsson diagram.
In hadron physics, the shape of a ``usual'' hadron is spherical.
The application of an external field, however, can distort the shape of hadrons and drastically modify their properties.
Such an intense field may appear in heavy-ion collision experiments: Noncentral collision between two charged nuclei can produce a strong magnetic field of which strength is estimated to reach $|eB| \sim m_\pi^2 \sim 0.02 \, \mathrm{GeV^2}$ at Relativistic Heavy Ion Collider (RHIC) and $|eB| \sim 15 m_\pi^2 \sim 0.3 \, \mathrm{GeV^2}$ at Large Hadron Collider (LHC) \cite{Kharzeev:2007jp,Skokov:2009qp}.
Furthermore, lattice QCD simulations provide useful method to study hadron properties in strong magnetic fields \cite{Bali:2011qj, Hidaka:2012mz, Luschevskaya:2014lga}.

For light mesons (with only up, down and strange quarks) in a magnetic field, many studies including $SU(3)$ lattice gauge theory \cite{Bali:2011qj, Hidaka:2012mz, Luschevskaya:2014lga} have been performed.
In particular, influence of the deformation of the QCD vacuum (so-called the magnetic catalysis) on hadron properties is one of the interesting topics.
Properties of hadrons with heavy quarks (charm or bottom) in a magnetic field were also investigated by various approaches including potential models \cite{Machado:2013rta,Alford:2013jva,Bonati:2015dka}, QCD sum rules \cite{Machado:2013yaa,Cho:2014exa,Cho:2014loa,Gubler:2015qok} and AdS/QCD \cite{Dudal:2014jfa}.
These works, however, have focused only to the ground-state hadrons. 

The main purpose of this paper is to investigate the deformation of the wave functions and corresponding energy levels for both ground and excited states of charmonia in a constant magnetic field, on the basis of the nonrelativistic quark model.
In such a model, there are two physical effects which lead to the change of hadronic properties: (i) {\it mixing between different spin states} by the $-\bm{\mu}_i \cdot \bm{B}$ term and (ii) {\it modification of quark kinetic energy} by the $\bm{B} \times \bm{r}$ term \footnote{
Recently, the anisotropy of a confinement potential deformed by a magnetic field was evaluated by lattice QCD \cite{Bonati:2014ksa} and its correction in the potential model for quarkonia was investigated in Ref.~\cite{Bonati:2015dka}}.

We note here that it is not technically easy to solve the two-body Schr\"{o}dinger equation with confining potential under strong magnetic field and to extract the anisotropic wave functions for ground and excited states simultaneously.
The previous attempt \cite{Alford:2013jva} by using the finite differential time domain (FDTD) method is so far limited only to the ground state. 

In this paper, we propose an approach based on a variational method, which we call cylindrical Gaussian expansion method (CGEM).
This method is an extension from the conventional Gaussian expansion method (GEM) \cite{Kamimura:1988zz,Hiyama:2003cu}.
The CGEM has the following nice properties: (i) It respects the symmetry of the Hamiltonian under constant magnetic field, (ii) it can deal fully with higher excited states and (iii) it reduces a computational cost substantially.

\section{The model}
We start with the same nonrelativistic two-body Hamiltonian as Ref.~\cite{Alford:2013jva},
\begin{equation}
H = \sum_{i=1}^2 \left[ \frac{1}{2m_i} \left( \bm{p}_i - q_i \bm{A} \right)^2 - \bm{\mu}_i \cdot \bm{B} +m_i \right] +V(r),
\end{equation}
where $m_i$, $q_i$ and $\bm{\mu}_i$ are the constituent quark mass, the quark electric charge and the quark magnetic moment, respectively.
For the vector potential, we choose the symmetric gauge $\bm{A}(\bm{r}_i) = \frac{1}{2} \bm{B} \times \bm{r}_i$.
We introduce the center of mass and relative coordinates, $\bm{R}=(m_1 \bm{r}_1 + m_2 \bm{r}_2)/M $ and $\bm{r} = \bm{r}_1 - \bm{r}_2$, where $M=m_1+m_2$ is the total mass of the two particles.
As a new conserved quantity, we define the pseudomomentum \cite{Avron:1978}: $\hat{\bm{K}} = \sum_{i=1}^2 \left[ \bm{p}_i +\frac{1}{2} q_i \bm{B} \times \bm{r}_i  \right]$ whose commutation relation is given by $[\hat{K}_i, \hat{K}_j] = -i(q_1+q_2) \epsilon_{ijk} B_k$.
When the system is charge neutral, the components $\hat{K}_i$ commute with each other. 
By using the pseudomomentum, the total wave function for the Hamiltonian can be factorized as follows: $\Phi(\bm{R},\bm{r}) = \exp \left[ i (\bm{K} -\frac{1}{2} q\bm{B} \times \bm{r}) \cdot \bm{R} \right] \Psi(\bm{r})$.
For a neutral system with $q_1=-q_2=q$, we can reduce the total Hamiltonian into
\begin{eqnarray}
H_\mathrm{rel} &=& \frac{\bm{K}^2}{2M} - \frac{q}{M}(\bm{K} \times \bm{B}) \cdot \bm{r} \nonumber\\
&& + \frac{q}{2} \left( \frac{1}{m_1} - \frac{1}{m_2} \right) \bm{B} \cdot( \bm{r} \times \bm{p}) \\
&& + \frac{\bm{p}^2}{2\mu} + \frac{q^2}{8\mu}( \bm{B} \times \bm{r})^2 + V(r) + \sum_{i=1}^2 [-\bm{\mu}_i \cdot \bm{B} + m_i], \nonumber
\end{eqnarray}
where $M=m_1+m_2$ and $\mu = m_1 m_2/M$.
If the orbital angular momentum along $\bm{B}$ is zero, we have $\bm{B} \cdot (\bm{r} \times \bm{p}) = 0$. 
When we put $\bm{B} =(0,0,B) $, $\bm{B} \times \bm{r} = B(-y,x,0)$ and $\bm{K} =(K_x,K_y,0) $, the Hamiltonian is reduced to 
\begin{eqnarray}
H_\mathrm{rel} &=& \frac{\bm{K}^2}{2M} -\frac{\nabla^2}{2\mu} + \frac{q^2 B^2}{8\mu} \rho^2 + \frac{qB}{4\mu} K_x y - \frac{qB}{4\mu} K_y x  \nonumber\\
&& + V(r) + \sum_{i=1}^2 [-\bm{\mu}_i \cdot \bm{B} + m_i]. \label{H_rel2}
\end{eqnarray}
In this work, we focus on only $\bm{K}=0$ in which the Hamiltonian maintains the rotational symmetry on the transverse $x$-$y$ plane and the reflection symmetry along the $z$-axis.

For the potential term, we choose the Cornell potential \cite{Eichten:1974af} with a spin-spin interaction:
\begin{eqnarray}
V(r) &=& \sigma r - \frac{A}{r} + \alpha (\bm{S}_1 \cdot \bm{S}_2)  e^{- \Lambda r^2} + C \nonumber\\
&=& \sigma \sqrt{\rho^2 + z^2} -\frac{A}{\sqrt{\rho^2 + z^2}} + \alpha (\bm{S}_1 \cdot \bm{S}_2) e^{- \Lambda (\rho^2 + z^2)} \nonumber\\
&& + C,
\end{eqnarray}
where $\bm{S}_1 \cdot \bm{S}_2 = -3/4$ and $1/4$ for the spin singlet and triplet, respectively.
Here, we adopted the conventional Gaussian form for the spin-spin interaction so that we have analytic expressions of the matrix elements.

Let us now consider the contribution of the spin mixing induced by the term $-\bm{\mu}_i \cdot \bm{B}$ \cite{Machado:2013rta,Alford:2013jva}, where the quark magnetic moment is given by $\bm{\mu}_i = g q_i \bm{S}_i/2m_i$ with the Land$\acute{\mathrm{e}}$ $g$-factor assumed to be $g=2$.
The eigenstates with the different spin quantum numbers (the singlet $| 00 \rangle$ and ``longitudinal" component $| 10 \rangle$ of the triplet) mix with each other through the off-diagonal matrix element: 
\begin{equation}
\langle 10 | (\bm{\mu}_1 + \bm{\mu}_2) \cdot \bm{B} |00 \rangle = -\frac{gqB}{4} \left( \frac{1}{m_1} + \frac{1}{m_2} \right). \label{mat_ele}
\end{equation}
On the other hand, ``transverse" components $| 1 \pm1 \rangle$ of the triplet cannot mix with other states since the couplings between the quark magnetic moments and the magnetic field are completely canceled between the quark and the antiquark.
To take into account the off-diagonal components of the matrix elements, Eq.~(\ref{mat_ele}), we have to solve a {\it coupled-channel} Schr\"{o}dinger equation.

\section{Cylindrical Gaussian expansion method (CGEM)}
In order to solve the two-body Schr\"{o}dinger equation, we use the GEM based on the Rayleigh-Ritz variational principal, which is a powerful tool in nuclear and atomic physics \cite{Kamimura:1988zz,Hiyama:2003cu}. 
According to the conventional GEM, in a system with the spherical symmetry, the radial wave function of a bound state can be expanded by Gaussian bases: $\sum_n C_n e^{-\alpha_n r^2}$ in the spherical coordinate $(r,\theta,\phi)$, where $C_n$ and $\alpha_n$ are the normalized expansion coefficient and the variational parameter (or range parameter), respectively.
On the other hand, the present system in the magnetic field does not have the spherical symmetry.
In this case, a wave function should be expanded individually by the Gaussian basis on the transverse plane, $e^{- \beta_n \rho^2}$, and that along the longitudinal axis, $e^{- \gamma_n z^2}$, in the cylindrical coordinate $(\rho,z,\phi)$.
Such bases are applied to atomic systems in a magnetic field (e.g. Ref.~\cite{Becken:2001}).

The trial wave function for $l_z=0$ is given as follows:
\begin{eqnarray}
&& \Psi (\rho,z,\phi) = \sum_{n=1}^N C_n \Phi_n (\rho,z,\phi), \\
&& \Phi_n (\rho,z,\phi) = N_n e^{- \beta_n \rho^2} e^{- \gamma_n z^2},
\end{eqnarray}
where $N$ and $C_n$ are the number of basis functions and the expansion coefficients.
$N_n$ is the normalization constant defined by $ \langle \Phi_n| \Phi_n \rangle =1$.
$\beta_n$ and $\gamma_n$ are Gaussian range parameters for $\rho$ and $z$ directions, respectively.
It is empirically known that the best set of range parameters, $\beta_1 \cdots \beta_N$ and $\gamma_1 \cdots \gamma_N$, are the geometric progressions: $\beta_n = 1/\rho_n^2$ with $\rho_n = \rho_1 b^{n-1}$ and $\gamma_n = 1/z_n^2$ with $z_n = z_1 c^{n-1}$.
Here, the four parameters, $\rho_1$, $\rho_N$, $z_1$ and $z_N$, will be optimized as the energy eigenvalue is minimized.
We checked the applicability of our numerical code by comparing our variational result with the analytic solution of the three-dimensional anisotropic harmonic oscillator \cite{Yoshida;prog}.

Here, we set the parameters of our constituent quark model.
For parameters in charmonium systems, the charm-quark kinetic mass $m_c=1.7840 \, \mathrm{GeV}$, the Coulomb parameter $A=0.713$ and the string tension $\sqrt{\sigma} = 0.402 \, \mathrm{GeV}$ are determined from the equal-time $Q\bar{Q}$ Bethe-Salpeter amplitude in lattice QCD \cite{Kawanai:2015tga}.
For the spin-dependent potential given as $ \alpha \exp{(-\Lambda r^2)}$, we adopt $\Lambda=1.020 \, \mathrm{GeV}^2$ obtained from lattice QCD \cite{Kawanai:2011jt} by fitting a charmonium potential.
The remaining parameters, $\alpha=0.4778 \, \rm{GeV}$ and $C=-0.5693 \, \rm{GeV}$, are chosen to reproduce the experimental values of the masses of the ground states of $\eta_c$ and $J/\psi$.
We checked that these parameters can reproduce well the masses of the ground and excited states of the charmonium.

\section{Results}

\begin{figure}[tb!]
     \centering
     \includegraphics[clip, width=1.0\columnwidth]{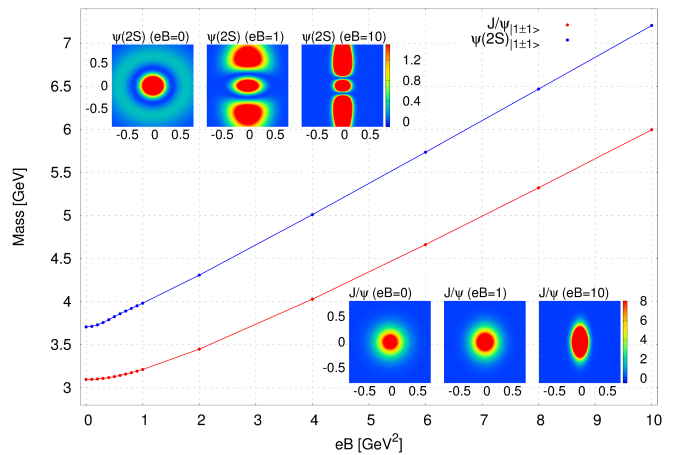}
    \caption{Transverse $J/\psi_{|1\pm 1 \rangle}$ in a magnetic field by calculation with single channel.
Wave functions $|\Psi|^2$ are plotted on plane with horizontal $x$ (or $y$) and vertical $z$ axes.
Units of numerical values on the spatial axes and color bar in the figures of the wave functions are $\mathrm{fm}$ and $\mathrm{fm}^3$, respectively.}
    \label{charmonia_T}
    \vspace{-5mm} 
\end{figure}

We investigate the region of the magnetic field up to $eB=10 \, \mathrm{GeV}^2$ to enable us to compare our results with future lattice QCD simulations, although the maximum of magnetic fields produced by heavy-ion collisions at LHC is $eB \sim 0.3 \, \mathrm{GeV}^2$.
Transverse $J/\psi$ is not mixed with other spin eigenstates, so that we can obtain the masses and the wave functions by solving the single-channel Schr\"{o}dinger equation.
The results are shown in Fig.~\ref{charmonia_T}.
In weak magnetic fields, the masses of the ground and excited states increase gradually.
Note that our result of the ground state in the weak magnetic fields agrees with that of Refs.~\cite{Alford:2013jva,Bonati:2015dka} obtained by the FDTD method.
In strong magnetic fields, the masses are linearly raised, which is consistent with the energy shift by the nonrelativistic Landau levels, $(n+1/2) |q|B/m_c$, of the {\it single} constituent quarks inside the hadron.
Thus, in this region, the single-particle picture for the mass shifts turns out to be a good approximation.
It is interesting that the mass shift of the excited $\psi(2S)$ state is larger than that of the ground state.
The reason is that the size of the wave function of an excited state is larger than that of the corresponding ground state, so that the expectation value of the $\rho^2$ term in Eq.~(\ref{H_rel2}) is also enhanced.

Moreover, the shape of the wave function is significantly modified.
In the ground state, the wave function on the $\rho$-plane is squeezed and the overall shape becomes cigar-shaped.
The wave function of the excited state with one spherical node in vacuum changes to the shape with one node along the $z$-direction.
This behavior implies that the excitation in the $\rho$-plane is removed by the $\rho^2$ term while that in the $z$-direction remains.

\begin{figure*}[t!]
    \centering
    \includegraphics[clip, width=15cm]{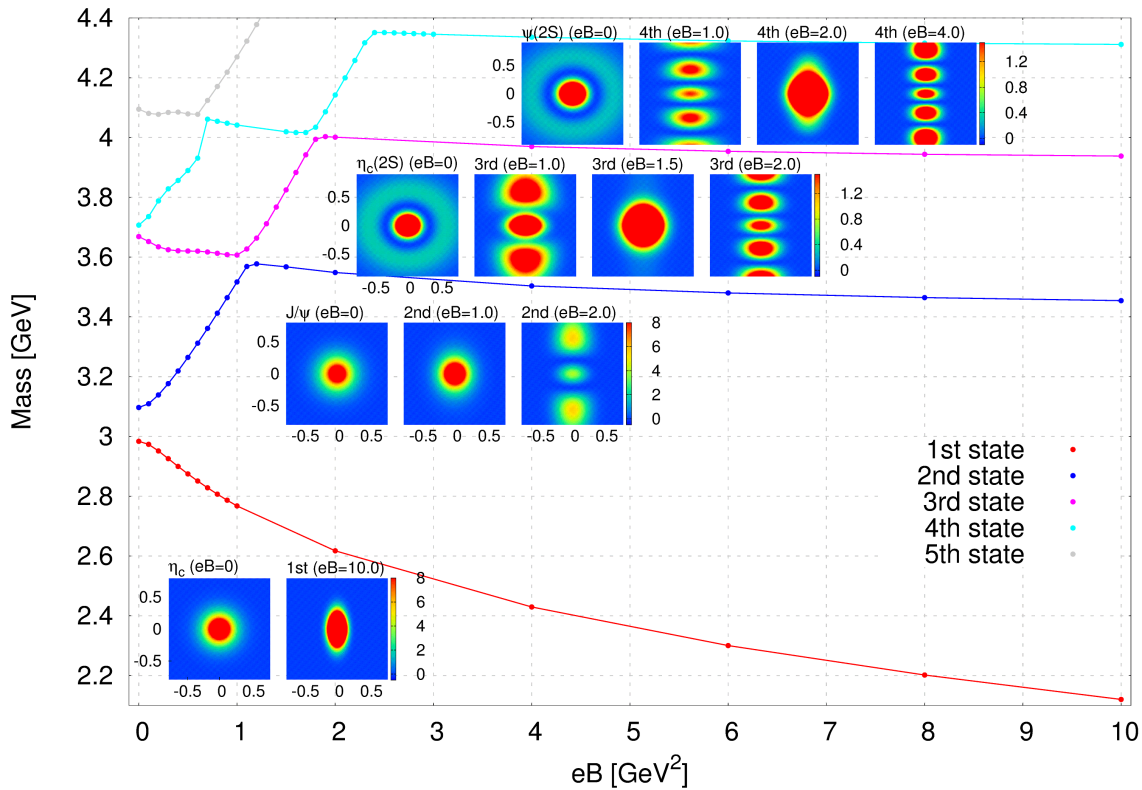}
    \caption{$\eta_c$ and longitudinal $J/\psi_{|10 \rangle}$ in a magnetic field by calculation with coupled channel.}
    \label{charmonia_L}
    \vspace{-5mm}
\end{figure*}

The results of $\eta_c$ and longitudinal $J/\psi$ are shown in Fig.~\ref{charmonia_L}.
These states are mixed with each other by $-\bm{\mu}_i \cdot \bm{B}$ term in Eq.~(\ref{H_rel2}), so that their masses change.
For the {\it first} state starting from $\eta_c(1S)$ at $eB=0$, the mixing partners include all excited states ($\psi(2S)$, $\psi(3S)$ ...) as well as the ground state $J/\psi$.
$\eta_c(1S)$ predominantly mixes with $J/\psi$ and its mass decreases gradually with increasing magnetic field.
The {\it second} state starting from $J/\psi$ at $eB=0$ is also mixed with $\eta_c(1S)$ in the weak magnetic fields.
As the magnetic field becomes larger, this state is contaminated by the excited states of $\eta_c$ and its mass approaches the third state like $\eta_c(2S)$ at $eB=1.1 \, \mathrm{GeV}^2$.
After that, the mass decreases slowly, where the wave function behaves as $2S$ state with one node.
As with the case of transverse $J/\psi$, the mass shifts of the first and second states in the weak magnetic fields are consistent with those of Refs.~\cite{Alford:2013jva,Bonati:2015dka}.

It is important to note that the mass shifts of the excited states in a weak magnetic field are more sensitive to $eB$ than that of the ground state.
The reason is that the hyperfine splitting between the spin partners in vacuum is narrower so that their mixing in a magnetic field becomes stronger.
Such a behavior can also be inferred from a simplified two-level model as discussed in Refs.~\cite{Alford:2013jva,Cho:2014exa,Cho:2014loa}.
For instance, the mass shifts of $\eta_c$, $J/\psi$, $\eta_c(2S)$ and $\psi(2S)$ at $eB=0.1 \, \mathrm{GeV}^2$ are $-9.9$, $+12.4$, $-16.6$ and $+28.8 \, \mathrm{MeV}$, respectively.
Thus the mixing in a weak magnetic field is consistent with the behavior of the two-level system while the behaviors in intermediate magnetic fields are more complicated.
The {\it third} state starting from $\eta_c(2S)$ approaches the second one like $J/\psi$ and, after that, the fourth one like $\eta_c(3S)$ at $eB=1.8 \, \mathrm{GeV}^2$.
Finally, similar behaviors are found in {\it fourth} state starting from $\psi(2S)$, which approaches the fifth state like $\eta_c(3S)$, the third state like $J/\psi$ and the fifth state like $\eta_c(4S)$.

\begin{figure}[b!]
\centering
     \includegraphics[clip, width=1.0\columnwidth]{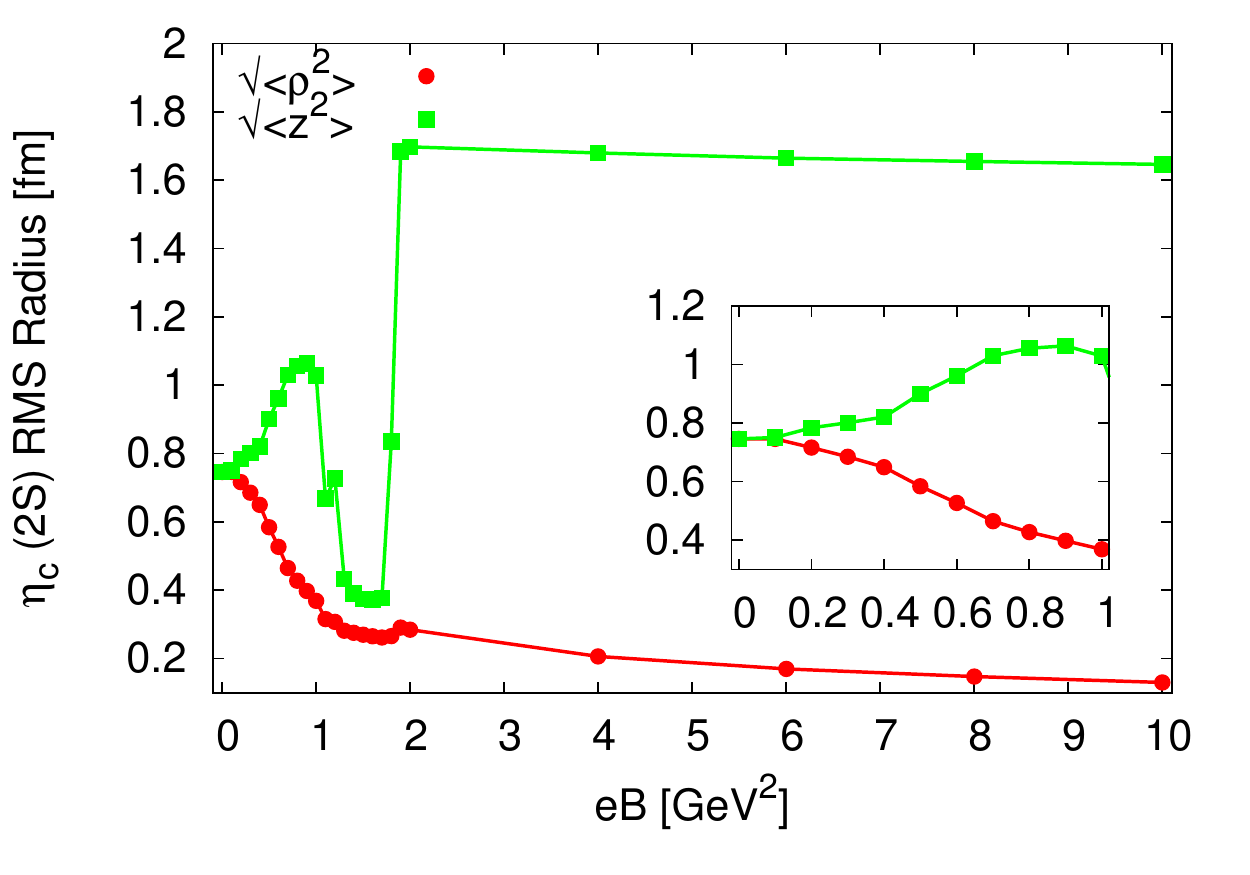}
    \caption{Root-mean square radii of the wave function of the third state in Fig.~\ref{charmonia_L}.}
    \label{eta_c2S_RMS}
\end{figure}

In Fig.~\ref{eta_c2S_RMS}, the root-mean square (RMS) radii of the wave function of the third state are shown.
We show radii defined by $\sqrt{\langle \rho^2 \rangle} \equiv \sqrt{(3/2) \langle \Psi | \rho^2 |\Psi \rangle}$ and $\sqrt{\langle z^2 \rangle} \equiv \sqrt{ 3 \langle \Psi | z^2 |\Psi \rangle}$ in the $\rho$- and $z$- directions, respectively.
These agree with the usual spherical RMS radius $\sqrt{\langle r^2 \rangle} $ in vacuum: $\sqrt{\langle r^2 \rangle}_{B=0} = \sqrt{\langle \rho^2 \rangle}_{B=0} = \sqrt{\langle z^2 \rangle}_{B=0}$.
In this figure, the decrease of $\sqrt{\langle \rho^2 \rangle}$ and the increase of $\sqrt{\langle z^2 \rangle}$ up to $eB=1.1 \, \mathrm{GeV}^2$ correspond to the shrinkage on the $\rho$-plane and the expansion in the $z$-direction for the wave function like $\eta_c(2S)$, respectively, as shown in Fig.~\ref{charmonia_L}.
Furthermore, the plateaus of $\sqrt{\langle z^2 \rangle}$ at $1.1<eB<1.8 \, \mathrm{GeV}^2$ and at $eB > 1.8 \, \mathrm{GeV}^2$ are corresponding to the wave functions like $J/\psi$ and $\eta_c(3S)$, respectively.

Let us discuss the bottomonia, such as $\eta_b(1S)$, $\Upsilon(1S)$, $\eta_b(2S)$, $\Upsilon(2S)$, $\eta_b(3S)$ and $\Upsilon(3S)$ below the threshold.
Their magnetic behaviors are similar to those of charmonia.
Since bottom quarks are heavier than charm quarks, the kinetic energies of the constituent quarks inside a hadron are suppressed and the sizes of the wave functions are smaller than those of charmonia.
Moreover, the electric charge of bottom quarks is $|q|=(1/3)|e|$ while that of charm quarks is $|q|=(2/3)|e|$, so that contributions of a magnetic field to bottomonia are also suppressed.
As a result, the mass shifts and wave functions of bottomonia are less sensitive to a magnetic field than those of charmonia.
Furthermore, the magnetic moment is suppressed by the mass and electric charge of bottom quarks, so that the mixing effect becomes also smaller. 
For instance, longitudinal $\Upsilon(1S)$ approaches $\eta_b(2S)$ at about $eB=5.5 \, \mathrm{GeV}^2$ \cite{Yoshida;prog}.

Our results in the weak magnetic fields provide various implications to the observables such as $\eta_c + \gamma \to J/\psi$ transition rate \cite{Yang:2011cz}, production cross section of quarkonia \cite{Machado:2013rta}, anisotropic production \cite{Guo:2015nsa} and quarkonium melting \cite{Marasinghe:2011bt,Dudal:2014jfa} in a magnetic field induced by heavy-ion collisions.
In particular, the dilepton spectra from vector quarkonia can shift and split between the transverse and longitudinal components \cite{Alford:2013jva,Filip:2013xza}.
As we discussed, the properties of the excited states are expected to be a sensitive probe of a magnetic field.
For instance, the mass shifts of $J/\psi_{|1\pm 1 \rangle}$, $\psi(2S)_{|1\pm 1 \rangle}$, $J/\psi_{|10 \rangle}$ and $\psi(2S)_{|10 \rangle}$ at $eB=0.3 \, \mathrm{GeV}^2$ are 12.4, 50.8, 79.7 and $121.5 \, \mathrm{MeV}$, respectively.
The mass splitting between the transverse and longitudinal components at $eB=0.3 \, \mathrm{GeV}^2$ are $67.3 \, \mathrm{MeV}$ and $70.7 \, \mathrm{MeV}$ for $J/\psi$ and $\psi(2S)$, respectively.
Furthermore, our predictions in the strong magnetic fields can be numerically confirmed by future lattice QCD simulations.

\section{Conclusion and outlook}
We investigated the static properties of quarkonia in homogeneous magnetic fields.
Our approach based on CGEM can deal fully with the coupled channel of the mixing between the spin-0 and spin-1 states, so that the resultant mass shifts include the mixing not only with the nearest spin partner but also with the other ground and excited states.
Moreover, the deformations of the wave functions in both the ground and excited states were discussed.
Some implications of our results to the experimental observables in RHIC and LHC were discussed.
To improve our results more quantitatively, we need to introduce relativistic corrections, an anisotropic confinement potential \cite{Miransky:2002rp,Bonati:2014ksa,Rougemont:2014efa,Bonati:2015dka} and modifications of the continuum threshold in the magnetic field.
It is an interesting future problem to extend our method into not only other neutral hadron systems such as $P$-wave mesons and baryons in a magnetic field, but also nonrelativistic bound states in atomic, molecular and nuclear few-body systems.
The investigations of the magnetic behaviors of such systems may help us to gain comprehensive insight into quantum many-body systems under strong magnetic field.

\begin{acknowledgments}
The authors thank Tetsuo Hatsuda and Makoto Oka for carefully reading our manuscript and giving us helpful advice.
We are grateful to Koichi Hattori and Shota Ohnishi for fruitful discussions.
K.S. was supported by Grant-in-Aid for JSPS Fellows from Japan Society for the Promotion of Science (JSPS) (No.26-8288).
\end{acknowledgments}

\bibliography{cigar}

\begin{thebibliography}{29}%
\makeatletter
\providecommand \@ifxundefined [1]{%
 \@ifx{#1\undefined}
}%
\providecommand \@ifnum [1]{%
 \ifnum #1\expandafter \@firstoftwo
 \else \expandafter \@secondoftwo
 \fi
}%
\providecommand \@ifx [1]{%
 \ifx #1\expandafter \@firstoftwo
 \else \expandafter \@secondoftwo
 \fi
}%
\providecommand \natexlab [1]{#1}%
\providecommand \enquote  [1]{``#1''}%
\providecommand \bibnamefont  [1]{#1}%
\providecommand \bibfnamefont [1]{#1}%
\providecommand \citenamefont [1]{#1}%
\providecommand \href@noop [0]{\@secondoftwo}%
\providecommand \href [0]{\begingroup \@sanitize@url \@href}%
\providecommand \@href[1]{\@@startlink{#1}\@@href}%
\providecommand \@@href[1]{\endgroup#1\@@endlink}%
\providecommand \@sanitize@url [0]{\catcode `\\12\catcode `\$12\catcode
  `\&12\catcode `\#12\catcode `\^12\catcode `\_12\catcode `\%12\relax}%
\providecommand \@@startlink[1]{}%
\providecommand \@@endlink[0]{}%
\providecommand \url  [0]{\begingroup\@sanitize@url \@url }%
\providecommand \@url [1]{\endgroup\@href {#1}{\urlprefix }}%
\providecommand \urlprefix  [0]{URL }%
\providecommand \Eprint [0]{\href }%
\providecommand \doibase [0]{http://dx.doi.org/}%
\providecommand \selectlanguage [0]{\@gobble}%
\providecommand \bibinfo  [0]{\@secondoftwo}%
\providecommand \bibfield  [0]{\@secondoftwo}%
\providecommand \translation [1]{[#1]}%
\providecommand \BibitemOpen [0]{}%
\providecommand \bibitemStop [0]{}%
\providecommand \bibitemNoStop [0]{.\EOS\space}%
\providecommand \EOS [0]{\spacefactor3000\relax}%
\providecommand \BibitemShut  [1]{\csname bibitem#1\endcsname}%
\let\auto@bib@innerbib\@empty
\bibitem [{\citenamefont {Kharzeev}\ \emph {et~al.}(2008)\citenamefont
  {Kharzeev}, \citenamefont {McLerran},\ and\ \citenamefont
  {Warringa}}]{Kharzeev:2007jp}%
  \BibitemOpen
  \bibfield  {author} {\bibinfo {author} {\bibfnamefont {D.~E.}\ \bibnamefont
  {Kharzeev}}, \bibinfo {author} {\bibfnamefont {L.~D.}\ \bibnamefont
  {McLerran}}, \ and\ \bibinfo {author} {\bibfnamefont {H.~J.}\ \bibnamefont
  {Warringa}},\ }\href {\doibase 10.1016/j.nuclphysa.2008.02.298} {\bibfield
  {journal} {\bibinfo  {journal} {Nucl. Phys.}\ }\textbf {\bibinfo {volume}
  {A803}},\ \bibinfo {pages} {227} (\bibinfo {year} {2008})},\ \Eprint
  {http://arxiv.org/abs/0711.0950} {arXiv:0711.0950 [hep-ph]} \BibitemShut
  {NoStop}%
\bibitem [{\citenamefont {Skokov}\ \emph {et~al.}(2009)\citenamefont {Skokov},
  \citenamefont {Illarionov},\ and\ \citenamefont {Toneev}}]{Skokov:2009qp}%
  \BibitemOpen
  \bibfield  {author} {\bibinfo {author} {\bibfnamefont {V.}~\bibnamefont
  {Skokov}}, \bibinfo {author} {\bibfnamefont {A.~{\relax Yu}.}\ \bibnamefont
  {Illarionov}}, \ and\ \bibinfo {author} {\bibfnamefont {V.}~\bibnamefont
  {Toneev}},\ }\href {\doibase 10.1142/S0217751X09047570} {\bibfield  {journal}
  {\bibinfo  {journal} {Int. J. Mod. Phys.}\ }\textbf {\bibinfo {volume}
  {A24}},\ \bibinfo {pages} {5925} (\bibinfo {year} {2009})},\ \Eprint
  {http://arxiv.org/abs/0907.1396} {arXiv:0907.1396 [nucl-th]} \BibitemShut
  {NoStop}%
\bibitem [{\citenamefont {Bali}\ \emph {et~al.}(2012)\citenamefont {Bali},
  \citenamefont {Bruckmann}, \citenamefont {Endr\"odi}, \citenamefont {Fodor},
  \citenamefont {Katz}, \citenamefont {Krieg}, \citenamefont {Schafer},\ and\
  \citenamefont {Szabo}}]{Bali:2011qj}%
  \BibitemOpen
  \bibfield  {author} {\bibinfo {author} {\bibfnamefont {G.~S.}\ \bibnamefont
  {Bali}}, \bibinfo {author} {\bibfnamefont {F.}~\bibnamefont {Bruckmann}},
  \bibinfo {author} {\bibfnamefont {G.}~\bibnamefont {Endr\"odi}}, \bibinfo
  {author} {\bibfnamefont {Z.}~\bibnamefont {Fodor}}, \bibinfo {author}
  {\bibfnamefont {S.~D.}\ \bibnamefont {Katz}}, \bibinfo {author}
  {\bibfnamefont {S.}~\bibnamefont {Krieg}}, \bibinfo {author} {\bibfnamefont
  {A.}~\bibnamefont {Schafer}}, \ and\ \bibinfo {author} {\bibfnamefont
  {K.~K.}\ \bibnamefont {Szabo}},\ }\href {\doibase 10.1007/JHEP02(2012)044}
  {\bibfield  {journal} {\bibinfo  {journal} {JHEP}\ }\textbf {\bibinfo
  {volume} {02}},\ \bibinfo {pages} {044} (\bibinfo {year} {2012})},\ \Eprint
  {http://arxiv.org/abs/1111.4956} {arXiv:1111.4956 [hep-lat]} \BibitemShut
  {NoStop}%
\bibitem [{\citenamefont {Hidaka}\ and\ \citenamefont
  {Yamamoto}(2013)}]{Hidaka:2012mz}%
  \BibitemOpen
  \bibfield  {author} {\bibinfo {author} {\bibfnamefont {Y.}~\bibnamefont
  {Hidaka}}\ and\ \bibinfo {author} {\bibfnamefont {A.}~\bibnamefont
  {Yamamoto}},\ }\href {\doibase 10.1103/PhysRevD.87.094502} {\bibfield
  {journal} {\bibinfo  {journal} {Phys.Rev.}\ }\textbf {\bibinfo {volume}
  {D87}},\ \bibinfo {pages} {094502} (\bibinfo {year} {2013})},\ \Eprint
  {http://arxiv.org/abs/1209.0007} {arXiv:1209.0007 [hep-ph]} \BibitemShut
  {NoStop}%
\bibitem [{\citenamefont {Luschevskaya}\ \emph {et~al.}(2015)\citenamefont
  {Luschevskaya}, \citenamefont {Solovjeva}, \citenamefont {Kochetkov},\ and\
  \citenamefont {Teryaev}}]{Luschevskaya:2014lga}%
  \BibitemOpen
  \bibfield  {author} {\bibinfo {author} {\bibfnamefont {E.~V.}\ \bibnamefont
  {Luschevskaya}}, \bibinfo {author} {\bibfnamefont {O.~E.}\ \bibnamefont
  {Solovjeva}}, \bibinfo {author} {\bibfnamefont {O.~A.}\ \bibnamefont
  {Kochetkov}}, \ and\ \bibinfo {author} {\bibfnamefont {O.~V.}\ \bibnamefont
  {Teryaev}},\ }\href {\doibase 10.1016/j.nuclphysb.2015.07.023} {\bibfield
  {journal} {\bibinfo  {journal} {Nucl. Phys.}\ }\textbf {\bibinfo {volume}
  {B898}},\ \bibinfo {pages} {627} (\bibinfo {year} {2015})},\ \Eprint
  {http://arxiv.org/abs/1411.4284} {arXiv:1411.4284 [hep-lat]} \BibitemShut
  {NoStop}%
\bibitem [{\citenamefont {Machado}\ \emph {et~al.}(2013)\citenamefont
  {Machado}, \citenamefont {Navarra}, \citenamefont {de~Oliveira},
  \citenamefont {Noronha},\ and\ \citenamefont {Strickland}}]{Machado:2013rta}%
  \BibitemOpen
  \bibfield  {author} {\bibinfo {author} {\bibfnamefont {C.~S.}\ \bibnamefont
  {Machado}}, \bibinfo {author} {\bibfnamefont {F.~S.}\ \bibnamefont
  {Navarra}}, \bibinfo {author} {\bibfnamefont {E.~G.}\ \bibnamefont
  {de~Oliveira}}, \bibinfo {author} {\bibfnamefont {J.}~\bibnamefont
  {Noronha}}, \ and\ \bibinfo {author} {\bibfnamefont {M.}~\bibnamefont
  {Strickland}},\ }\href {\doibase 10.1103/PhysRevD.88.034009} {\bibfield
  {journal} {\bibinfo  {journal} {Phys. Rev.}\ }\textbf {\bibinfo {volume}
  {D88}},\ \bibinfo {pages} {034009} (\bibinfo {year} {2013})},\ \Eprint
  {http://arxiv.org/abs/1305.3308} {arXiv:1305.3308 [hep-ph]} \BibitemShut
  {NoStop}%
\bibitem [{\citenamefont {Alford}\ and\ \citenamefont
  {Strickland}(2013)}]{Alford:2013jva}%
  \BibitemOpen
  \bibfield  {author} {\bibinfo {author} {\bibfnamefont {J.}~\bibnamefont
  {Alford}}\ and\ \bibinfo {author} {\bibfnamefont {M.}~\bibnamefont
  {Strickland}},\ }\href {\doibase 10.1103/PhysRevD.88.105017} {\bibfield
  {journal} {\bibinfo  {journal} {Phys.Rev.}\ }\textbf {\bibinfo {volume}
  {D88}},\ \bibinfo {pages} {105017} (\bibinfo {year} {2013})},\ \Eprint
  {http://arxiv.org/abs/1309.3003} {arXiv:1309.3003 [hep-ph]} \BibitemShut
  {NoStop}%
\bibitem [{\citenamefont {Bonati}\ \emph {et~al.}(2015)\citenamefont {Bonati},
  \citenamefont {D'Elia},\ and\ \citenamefont {Rucci}}]{Bonati:2015dka}%
  \BibitemOpen
  \bibfield  {author} {\bibinfo {author} {\bibfnamefont {C.}~\bibnamefont
  {Bonati}}, \bibinfo {author} {\bibfnamefont {M.}~\bibnamefont {D'Elia}}, \
  and\ \bibinfo {author} {\bibfnamefont {A.}~\bibnamefont {Rucci}},\ }\href
  {\doibase 10.1103/PhysRevD.92.054014} {\bibfield  {journal} {\bibinfo
  {journal} {Phys. Rev.}\ }\textbf {\bibinfo {volume} {D92}},\ \bibinfo {pages}
  {054014} (\bibinfo {year} {2015})},\ \Eprint
  {http://arxiv.org/abs/1506.07890} {arXiv:1506.07890 [hep-ph]} \BibitemShut
  {NoStop}%
\bibitem [{\citenamefont {Machado}\ \emph {et~al.}(2014)\citenamefont
  {Machado}, \citenamefont {Matheus}, \citenamefont {Finazzo},\ and\
  \citenamefont {Noronha}}]{Machado:2013yaa}%
  \BibitemOpen
  \bibfield  {author} {\bibinfo {author} {\bibfnamefont {C.~S.}\ \bibnamefont
  {Machado}}, \bibinfo {author} {\bibfnamefont {R.~D.}\ \bibnamefont
  {Matheus}}, \bibinfo {author} {\bibfnamefont {S.~I.}\ \bibnamefont
  {Finazzo}}, \ and\ \bibinfo {author} {\bibfnamefont {J.}~\bibnamefont
  {Noronha}},\ }\href {\doibase 10.1103/PhysRevD.89.074027} {\bibfield
  {journal} {\bibinfo  {journal} {Phys. Rev.}\ }\textbf {\bibinfo {volume}
  {D89}},\ \bibinfo {pages} {074027} (\bibinfo {year} {2014})},\ \Eprint
  {http://arxiv.org/abs/1307.1797} {arXiv:1307.1797 [hep-ph]} \BibitemShut
  {NoStop}%
\bibitem [{\citenamefont {Cho}\ \emph {et~al.}(2014)\citenamefont {Cho},
  \citenamefont {Hattori}, \citenamefont {Lee}, \citenamefont {Morita},\ and\
  \citenamefont {Ozaki}}]{Cho:2014exa}%
  \BibitemOpen
  \bibfield  {author} {\bibinfo {author} {\bibfnamefont {S.}~\bibnamefont
  {Cho}}, \bibinfo {author} {\bibfnamefont {K.}~\bibnamefont {Hattori}},
  \bibinfo {author} {\bibfnamefont {S.~H.}\ \bibnamefont {Lee}}, \bibinfo
  {author} {\bibfnamefont {K.}~\bibnamefont {Morita}}, \ and\ \bibinfo {author}
  {\bibfnamefont {S.}~\bibnamefont {Ozaki}},\ }\href {\doibase
  10.1103/PhysRevLett.113.172301} {\bibfield  {journal} {\bibinfo  {journal}
  {Phys. Rev. Lett.}\ }\textbf {\bibinfo {volume} {113}},\ \bibinfo {pages}
  {172301} (\bibinfo {year} {2014})},\ \Eprint {http://arxiv.org/abs/1406.4586}
  {arXiv:1406.4586 [hep-ph]} \BibitemShut {NoStop}%
\bibitem [{\citenamefont {Cho}\ \emph {et~al.}(2015)\citenamefont {Cho},
  \citenamefont {Hattori}, \citenamefont {Lee}, \citenamefont {Morita},\ and\
  \citenamefont {Ozaki}}]{Cho:2014loa}%
  \BibitemOpen
  \bibfield  {author} {\bibinfo {author} {\bibfnamefont {S.}~\bibnamefont
  {Cho}}, \bibinfo {author} {\bibfnamefont {K.}~\bibnamefont {Hattori}},
  \bibinfo {author} {\bibfnamefont {S.~H.}\ \bibnamefont {Lee}}, \bibinfo
  {author} {\bibfnamefont {K.}~\bibnamefont {Morita}}, \ and\ \bibinfo {author}
  {\bibfnamefont {S.}~\bibnamefont {Ozaki}},\ }\href {\doibase
  10.1103/PhysRevD.91.045025} {\bibfield  {journal} {\bibinfo  {journal} {Phys.
  Rev.}\ }\textbf {\bibinfo {volume} {D91}},\ \bibinfo {pages} {045025}
  (\bibinfo {year} {2015})},\ \Eprint {http://arxiv.org/abs/1411.7675}
  {arXiv:1411.7675 [hep-ph]} \BibitemShut {NoStop}%
\bibitem [{\citenamefont {Gubler}\ \emph {et~al.}(2016)\citenamefont {Gubler},
  \citenamefont {Hattori}, \citenamefont {Lee}, \citenamefont {Oka},
  \citenamefont {Ozaki},\ and\ \citenamefont {Suzuki}}]{Gubler:2015qok}%
  \BibitemOpen
  \bibfield  {author} {\bibinfo {author} {\bibfnamefont {P.}~\bibnamefont
  {Gubler}}, \bibinfo {author} {\bibfnamefont {K.}~\bibnamefont {Hattori}},
  \bibinfo {author} {\bibfnamefont {S.~H.}\ \bibnamefont {Lee}}, \bibinfo
  {author} {\bibfnamefont {M.}~\bibnamefont {Oka}}, \bibinfo {author}
  {\bibfnamefont {S.}~\bibnamefont {Ozaki}}, \ and\ \bibinfo {author}
  {\bibfnamefont {K.}~\bibnamefont {Suzuki}},\ }\href {\doibase
  10.1103/PhysRevD.93.054026} {\bibfield  {journal} {\bibinfo  {journal} {Phys.
  Rev.}\ }\textbf {\bibinfo {volume} {D93}},\ \bibinfo {pages} {054026}
  (\bibinfo {year} {2016})},\ \Eprint {http://arxiv.org/abs/1512.08864}
  {arXiv:1512.08864 [hep-ph]} \BibitemShut {NoStop}%
\bibitem [{\citenamefont {Dudal}\ and\ \citenamefont
  {Mertens}(2015)}]{Dudal:2014jfa}%
  \BibitemOpen
  \bibfield  {author} {\bibinfo {author} {\bibfnamefont {D.}~\bibnamefont
  {Dudal}}\ and\ \bibinfo {author} {\bibfnamefont {T.~G.}\ \bibnamefont
  {Mertens}},\ }\href {\doibase 10.1103/PhysRevD.91.086002} {\bibfield
  {journal} {\bibinfo  {journal} {Phys. Rev.}\ }\textbf {\bibinfo {volume}
  {D91}},\ \bibinfo {pages} {086002} (\bibinfo {year} {2015})},\ \Eprint
  {http://arxiv.org/abs/1410.3297} {arXiv:1410.3297 [hep-th]} \BibitemShut
  {NoStop}%
\bibitem [{Note1()}]{Note1}%
  \BibitemOpen
  \bibinfo {note} {Recently, the anisotropy of a confinement potential deformed
  by a magnetic field was evaluated by lattice QCD \cite {Bonati:2014ksa} and
  its correction in the potential model for quarkonia was investigated in
  Ref.~\cite {Bonati:2015dka}}\BibitemShut {NoStop}%
\bibitem [{\citenamefont {Kamimura}(1988)}]{Kamimura:1988zz}%
  \BibitemOpen
  \bibfield  {author} {\bibinfo {author} {\bibfnamefont {M.}~\bibnamefont
  {Kamimura}},\ }\href {\doibase 10.1103/PhysRevA.38.621} {\bibfield  {journal}
  {\bibinfo  {journal} {Phys. Rev.}\ }\textbf {\bibinfo {volume} {A38}},\
  \bibinfo {pages} {621} (\bibinfo {year} {1988})}\BibitemShut {NoStop}%
\bibitem [{\citenamefont {Hiyama}\ \emph {et~al.}(2003)\citenamefont {Hiyama},
  \citenamefont {Kino},\ and\ \citenamefont {Kamimura}}]{Hiyama:2003cu}%
  \BibitemOpen
  \bibfield  {author} {\bibinfo {author} {\bibfnamefont {E.}~\bibnamefont
  {Hiyama}}, \bibinfo {author} {\bibfnamefont {Y.}~\bibnamefont {Kino}}, \ and\
  \bibinfo {author} {\bibfnamefont {M.}~\bibnamefont {Kamimura}},\ }\href
  {\doibase 10.1016/S0146-6410(03)90015-9} {\bibfield  {journal} {\bibinfo
  {journal} {Prog. Part. Nucl. Phys.}\ }\textbf {\bibinfo {volume} {51}},\
  \bibinfo {pages} {223} (\bibinfo {year} {2003})}\BibitemShut {NoStop}%
\bibitem [{\citenamefont {Avron}\ \emph {et~al.}(1978)\citenamefont {Avron},
  \citenamefont {Herbst},\ and\ \citenamefont {Simon}}]{Avron:1978}%
  \BibitemOpen
  \bibfield  {author} {\bibinfo {author} {\bibfnamefont {J.~E.}\ \bibnamefont
  {Avron}}, \bibinfo {author} {\bibfnamefont {I.~W.}\ \bibnamefont {Herbst}}, \
  and\ \bibinfo {author} {\bibfnamefont {B.}~\bibnamefont {Simon}},\ }\href
  {\doibase 10.1016/0003-4916(78)90276-2} {\bibfield  {journal} {\bibinfo
  {journal} {Ann. Phys. (N.Y.)}\ }\textbf {\bibinfo {volume} {114}},\ \bibinfo
  {pages} {431} (\bibinfo {year} {1978})}\BibitemShut {NoStop}%
\bibitem [{\citenamefont {Eichten}\ \emph {et~al.}(1975)\citenamefont
  {Eichten}, \citenamefont {Gottfried}, \citenamefont {Kinoshita},
  \citenamefont {Kogut}, \citenamefont {Lane},\ and\ \citenamefont
  {Yan}}]{Eichten:1974af}%
  \BibitemOpen
  \bibfield  {author} {\bibinfo {author} {\bibfnamefont {E.}~\bibnamefont
  {Eichten}}, \bibinfo {author} {\bibfnamefont {K.}~\bibnamefont {Gottfried}},
  \bibinfo {author} {\bibfnamefont {T.}~\bibnamefont {Kinoshita}}, \bibinfo
  {author} {\bibfnamefont {J.~B.}\ \bibnamefont {Kogut}}, \bibinfo {author}
  {\bibfnamefont {K.~D.}\ \bibnamefont {Lane}}, \ and\ \bibinfo {author}
  {\bibfnamefont {T.-M.}\ \bibnamefont {Yan}},\ }\href {\doibase
  10.1103/PhysRevLett.34.369} {\bibfield  {journal} {\bibinfo  {journal} {Phys.
  Rev. Lett.}\ }\textbf {\bibinfo {volume} {34}},\ \bibinfo {pages} {369}
  (\bibinfo {year} {1975})},\ \bibinfo {note} {[Erratum: Phys. Rev.
  Lett.36,1276(1976)]}\BibitemShut {NoStop}%
\bibitem [{\citenamefont {Becken}\ and\ \citenamefont
  {Schmelcher}(2001)}]{Becken:2001}%
  \BibitemOpen
  \bibfield  {author} {\bibinfo {author} {\bibfnamefont {W.}~\bibnamefont
  {Becken}}\ and\ \bibinfo {author} {\bibfnamefont {P.}~\bibnamefont
  {Schmelcher}},\ }\href {\doibase 10.1103/PhysRevA.63.053412} {\bibfield
  {journal} {\bibinfo  {journal} {Phys. Rev.}\ }\textbf {\bibinfo {volume}
  {A63}},\ \bibinfo {pages} {053412} (\bibinfo {year} {2001})}\BibitemShut
  {NoStop}%
\bibitem [{\citenamefont {Yoshida}\ and\ \citenamefont
  {Suzuki}()}]{Yoshida;prog}%
  \BibitemOpen
  \bibfield  {author} {\bibinfo {author} {\bibfnamefont {T.}~\bibnamefont
  {Yoshida}}\ and\ \bibinfo {author} {\bibfnamefont {K.}~\bibnamefont
  {Suzuki}},\ }\href@noop {} {\bibinfo  {journal} {(to be published)}\
  }\BibitemShut {NoStop}%
\bibitem [{\citenamefont {Kawanai}\ and\ \citenamefont
  {Sasaki}(2015)}]{Kawanai:2015tga}%
  \BibitemOpen
\bibfield  {journal} {  }\bibfield  {author} {\bibinfo {author} {\bibfnamefont
  {T.}~\bibnamefont {Kawanai}}\ and\ \bibinfo {author} {\bibfnamefont
  {S.}~\bibnamefont {Sasaki}},\ }\href {\doibase 10.1103/PhysRevD.92.094503}
  {\bibfield  {journal} {\bibinfo  {journal} {Phys. Rev.}\ }\textbf {\bibinfo
  {volume} {D92}},\ \bibinfo {pages} {094503} (\bibinfo {year} {2015})},\
  \Eprint {http://arxiv.org/abs/1508.02178} {arXiv:1508.02178 [hep-lat]}
  \BibitemShut {NoStop}%
\bibitem [{\citenamefont {Kawanai}\ and\ \citenamefont
  {Sasaki}(2012)}]{Kawanai:2011jt}%
  \BibitemOpen
  \bibfield  {author} {\bibinfo {author} {\bibfnamefont {T.}~\bibnamefont
  {Kawanai}}\ and\ \bibinfo {author} {\bibfnamefont {S.}~\bibnamefont
  {Sasaki}},\ }\href {\doibase 10.1103/PhysRevD.85.091503} {\bibfield
  {journal} {\bibinfo  {journal} {Phys. Rev.}\ }\textbf {\bibinfo {volume}
  {D85}},\ \bibinfo {pages} {091503} (\bibinfo {year} {2012})},\ \Eprint
  {http://arxiv.org/abs/1110.0888} {arXiv:1110.0888 [hep-lat]} \BibitemShut
  {NoStop}%
\bibitem [{\citenamefont {Yang}\ and\ \citenamefont {M{\"
  u}ller}(2012)}]{Yang:2011cz}%
  \BibitemOpen
  \bibfield  {author} {\bibinfo {author} {\bibfnamefont {D.-L.}\ \bibnamefont
  {Yang}}\ and\ \bibinfo {author} {\bibfnamefont {B.}~\bibnamefont {M{\"
  u}ller}},\ }\href {\doibase 10.1088/0954-3899/39/1/015007} {\bibfield
  {journal} {\bibinfo  {journal} {J. Phys.}\ }\textbf {\bibinfo {volume}
  {G39}},\ \bibinfo {pages} {015007} (\bibinfo {year} {2012})},\ \Eprint
  {http://arxiv.org/abs/1108.2525} {arXiv:1108.2525 [hep-ph]} \BibitemShut
  {NoStop}%
\bibitem [{\citenamefont {Guo}\ \emph {et~al.}(2015)\citenamefont {Guo},
  \citenamefont {Shi}, \citenamefont {Xu}, \citenamefont {Xu},\ and\
  \citenamefont {Zhuang}}]{Guo:2015nsa}%
  \BibitemOpen
  \bibfield  {author} {\bibinfo {author} {\bibfnamefont {X.}~\bibnamefont
  {Guo}}, \bibinfo {author} {\bibfnamefont {S.}~\bibnamefont {Shi}}, \bibinfo
  {author} {\bibfnamefont {N.}~\bibnamefont {Xu}}, \bibinfo {author}
  {\bibfnamefont {Z.}~\bibnamefont {Xu}}, \ and\ \bibinfo {author}
  {\bibfnamefont {P.}~\bibnamefont {Zhuang}},\ }\href {\doibase
  10.1016/j.physletb.2015.10.038} {\bibfield  {journal} {\bibinfo  {journal}
  {Phys. Lett.}\ }\textbf {\bibinfo {volume} {B751}},\ \bibinfo {pages} {215}
  (\bibinfo {year} {2015})},\ \Eprint {http://arxiv.org/abs/1502.04407}
  {arXiv:1502.04407 [hep-ph]} \BibitemShut {NoStop}%
\bibitem [{\citenamefont {Marasinghe}\ and\ \citenamefont
  {Tuchin}(2011)}]{Marasinghe:2011bt}%
  \BibitemOpen
  \bibfield  {author} {\bibinfo {author} {\bibfnamefont {K.}~\bibnamefont
  {Marasinghe}}\ and\ \bibinfo {author} {\bibfnamefont {K.}~\bibnamefont
  {Tuchin}},\ }\href {\doibase 10.1103/PhysRevC.84.044908} {\bibfield
  {journal} {\bibinfo  {journal} {Phys. Rev.}\ }\textbf {\bibinfo {volume}
  {C84}},\ \bibinfo {pages} {044908} (\bibinfo {year} {2011})},\ \Eprint
  {http://arxiv.org/abs/1103.1329} {arXiv:1103.1329 [hep-ph]} \BibitemShut
  {NoStop}%
\bibitem [{\citenamefont {Filip}(2013)}]{Filip:2013xza}%
  \BibitemOpen
  \bibfield  {author} {\bibinfo {author} {\bibfnamefont {P.}~\bibnamefont
  {Filip}},\ }\href@noop {} {\bibfield  {journal} {\bibinfo  {journal} {PoS}\
  }\textbf {\bibinfo {volume} {CPOD2013}},\ \bibinfo {pages} {035} (\bibinfo
  {year} {2013})}\BibitemShut {NoStop}%
\bibitem [{\citenamefont {Miransky}\ and\ \citenamefont
  {Shovkovy}(2002)}]{Miransky:2002rp}%
  \BibitemOpen
  \bibfield  {author} {\bibinfo {author} {\bibfnamefont {V.~A.}\ \bibnamefont
  {Miransky}}\ and\ \bibinfo {author} {\bibfnamefont {I.~A.}\ \bibnamefont
  {Shovkovy}},\ }\href {\doibase 10.1103/PhysRevD.66.045006} {\bibfield
  {journal} {\bibinfo  {journal} {Phys. Rev.}\ }\textbf {\bibinfo {volume}
  {D66}},\ \bibinfo {pages} {045006} (\bibinfo {year} {2002})},\ \Eprint
  {http://arxiv.org/abs/hep-ph/0205348} {arXiv:hep-ph/0205348 [hep-ph]}
  \BibitemShut {NoStop}%
\bibitem [{\citenamefont {Bonati}\ \emph {et~al.}(2014)\citenamefont {Bonati},
  \citenamefont {D'Elia}, \citenamefont {Mariti}, \citenamefont {Mesiti},
  \citenamefont {Negro},\ and\ \citenamefont {Sanfilippo}}]{Bonati:2014ksa}%
  \BibitemOpen
  \bibfield  {author} {\bibinfo {author} {\bibfnamefont {C.}~\bibnamefont
  {Bonati}}, \bibinfo {author} {\bibfnamefont {M.}~\bibnamefont {D'Elia}},
  \bibinfo {author} {\bibfnamefont {M.}~\bibnamefont {Mariti}}, \bibinfo
  {author} {\bibfnamefont {M.}~\bibnamefont {Mesiti}}, \bibinfo {author}
  {\bibfnamefont {F.}~\bibnamefont {Negro}}, \ and\ \bibinfo {author}
  {\bibfnamefont {F.}~\bibnamefont {Sanfilippo}},\ }\href {\doibase
  10.1103/PhysRevD.89.114502} {\bibfield  {journal} {\bibinfo  {journal} {Phys.
  Rev.}\ }\textbf {\bibinfo {volume} {D89}},\ \bibinfo {pages} {114502}
  (\bibinfo {year} {2014})},\ \Eprint {http://arxiv.org/abs/1403.6094}
  {arXiv:1403.6094 [hep-lat]} \BibitemShut {NoStop}%
\bibitem [{\citenamefont {Rougemont}\ \emph {et~al.}(2015)\citenamefont
  {Rougemont}, \citenamefont {Critelli},\ and\ \citenamefont
  {Noronha}}]{Rougemont:2014efa}%
  \BibitemOpen
  \bibfield  {author} {\bibinfo {author} {\bibfnamefont {R.}~\bibnamefont
  {Rougemont}}, \bibinfo {author} {\bibfnamefont {R.}~\bibnamefont {Critelli}},
  \ and\ \bibinfo {author} {\bibfnamefont {J.}~\bibnamefont {Noronha}},\ }\href
  {\doibase 10.1103/PhysRevD.91.066001} {\bibfield  {journal} {\bibinfo
  {journal} {Phys. Rev.}\ }\textbf {\bibinfo {volume} {D91}},\ \bibinfo {pages}
  {066001} (\bibinfo {year} {2015})},\ \Eprint {http://arxiv.org/abs/1409.0556}
  {arXiv:1409.0556 [hep-th]} \BibitemShut {NoStop}%
\end{thebibliography}%

\end{document}